\documentclass[12pt]{article}
\pdfoutput=1
\usepackage{amsbsy}
\usepackage{amsfonts}
\usepackage{amsmath,amssymb}
\usepackage{bbold}
\usepackage{pdfpages}

\newcommand{\sect}[1]{\setcounter{equation}{0}\section{#1}}

\newcommand{\eq}{\begin{equation}}
\newcommand{\eqa}{\begin{eqnarray}}  
\newcommand{\en}{\end{equation}}
\newcommand{\ena}{\end{eqnarray}}
\newcommand{\enn}{\nonumber \end{equation}}

\def\sk{\vskip .4cm}
\def\noi{\noindent}
\def\om{\omega}
\def\al{\alpha}
\def\be{\beta}
\def\ga{\gamma}

\let \si\sigma
\let \part\partial

\def\unquarto{{1 \over 4}}

\def\unmezzo{{1 \over 2}}
\def\epsi{\varepsilon}
\def\we{\wedge}

\def\de{\delta}

\def\Gt{{\tilde G}}

\def\part{\partial}

\def\sk{\vskip .4cm}

\def\noi{\noindent}

\def\pdyN{{\partial \over \partial y^N}}

\def\X0{X^0}

\def\om{\omega}

\def\al{\alpha}
\def\ga{\gamma}

\def\unquarto{{1 \over 4}}
\def\unmezzo{{1 \over 2}}
\def\epsi{\varepsilon}

\def\psib{{\bar \psi}}

\def\rhob{\bar\rho}

\def\we{\wedge}

\def\de{\delta}
\def\CABC{{C^A}_{BC}}
\def\CCAB{{C^C}_{AB}}

\def\Gt{{\tilde G}}

\def\Dcal{{\cal D}}
\def\Rcal{{\cal R}}

\def\square{{\,\lower0.9pt\vbox{\hrule \hbox{\vrule height 0.2 cm
\hskip 0.2 cm \vrule height 0.2 cm}\hrule}\,}}

\def\epsilonbar{{\bar \epsilon}}
\def\thetabar{{\bar \theta}}

\def\Gtilde{\tilde G}

\def\psibar{\bar \psi}
\def\epsilonbar{\bar \epsilon}
\def\chibar{\bar \chi}
\def\rhobar{\bar \rho}

\def\Om{\Omega}
\def\Qbar{\overline{Q}}

\begin{document}

\begin{titlepage}
\vskip 2em
\begin{center}
{\Large \bf Group manifold approach to supergravity} \\[3em]

\vskip 0.5cm

{\bf
Leonardo Castellani}
\medskip

\vskip 0.5cm

{\sl Dipartimento di Scienze e Innovazione Tecnologica
\\Universit\`a del Piemonte Orientale, viale T. Michel 11, 15121 Alessandria, Italy\\ [.5em] INFN, Sezione di 
Torino, via P. Giuria 1, 10125 Torino, Italy\\ [.5em]
Arnold-Regge Center, via P. Giuria 1, 10125 Torino, Italy
}\\ [4em]
\end{center}

\begin{abstract}
\sk

We present a short review of the group-geometric approach to supergravity theories, from the point of view
of recent developments. The central idea is the unification of usual diffeomorphisms, gauge symmetries and supersymmetries into superdiffeomorphisms in a supergroup manifold.  The example of $N=1$ supergravity in $d=4$ is discussed in detail, and used to illustrate all the steps in the construction of a group manifold action.  In the Appendices we summarize basic notions of group manifold geometry, and of integration on supermanifolds. 
\end{abstract}

\vskip 4cm\sk\sk
\noi {\small Invited chapter for the ``Handbook of Quantum Gravity", Eds. C. Bambi, L. Modesto and I.L. Shapiro, Springer, expected in 2023.}
\vskip 2cm
 \noi \hrule \vskip .2cm \noi {\small
leonardo.castellani@uniupo.it}

\end{titlepage}

\newpage
\setcounter{page}{1}

\tableofcontents

\sect{Introduction}

Fundamental interactions are described by field theories with local invariances: the actions
that govern their dynamics are invariant under field transformations involving parameters that are (arbitrary) functions of spacetime. This holds true both for gravity and gauge theories, where the local symmetries are general coordinate, and gauge transformations, respectively.

The essential difference between these two types of local transformations, in their infinitesimal versions, is that diffeomorphisms always contain a derivative of the field,  which is absent in gauge transformations. As well known, this is due to the fact that general coordinate transformations relate fields at different spacetime points, whereas gauge transformations relate fields at the same spacetime point.

Nonetheless, it is possible to give a unified description of diffeomorphisms and gauge transformations.
This we achieve in a group geometrical framework. 

The main idea is to consider as basic fields of the theory the {\sl components of the vielbein} one-form $\sigma^A=\sigma(z)^A_{~\Lambda} dz^\Lambda$ on the manifold of a Lie (super)group $G$,  {\small {\it A}} being an index in the $G$ Lie (super)algebra, and $z^\Lambda$ the coordinates of the group manifold.
 This vielbein satisfies the Cartan-Maurer (CM) equations
 \eq
d \sigma^A + {1 \over 2} C^A_{BC} ~\sigma^B \wedge \sigma^C =0 \label{CM}
\en
where $C^A_{BC}$ are the structure constants of the $G$ Lie algebra. A brief account of group manifold geometry is given in Appendix A.

The $G$ vielbein $\sigma^A (z)$  has a fixed dependence on the coordinates $z$, and therefore cannot
be a dynamical object. We must consider a ``soft" group manifold, 
diffeomorphic to $G$ and denoted by ${\tilde G}$,
with a vielbein $\sigma^A$ not satisfying anymore the CM equations. The amount of deformation from the original ``rigid" group manifold is measured by the {\sl curvature} two-form:
\eq
R^A \equiv d \sigma^A + {1 \over 2} C^A_{BC} ~\sigma^B \wedge \sigma^C  \label{Gcurvature}
\en
Tangent vectors
on $\Gtilde$, dual to the vielbein $\sigma^A$, are denoted by $t_B$, so that $\sigma^A (t_B)=\delta^A_B$.
\sk
Diffeomorphisms along tangent vectors $\epsi=\epsi^A t_A$ on $\Gtilde$ are generated by the Lie derivative $\ell_\epsi$.  When applied to the $\Gtilde$ vielbein, the variation under diffeomorphisms takes the  form:
\eq
\ell_\epsi \sigma^A = d \epsi^A + C^A_{BC} \sigma^B \epsi^C + \iota_\epsi R^A \label{Lieder}
\en
where $\iota_\epsi$ is the contraction operator, see Appendix A.  On the right-hand side one recognizes the $G$-covariant derivative of the infinitesimal parameter 
 $\epsi^A$ plus a curvature term. When the curvature term vanishes, i.e. when 
 $\iota_\epsi R^A=0$, the diffeomorphism
 takes the form of a {\it gauge transformation},  and the
 curvature is said to be {\it horizontal}
 along the $t_A$'s entering the sum in $\epsi=\epsi^A t_A$.
 Thus in group manifold geometry {\it gauge transformations} can be interpreted as {\it particular
 diffeomorphisms}, along the directions on which the curvatures are horizontal. 
 
 This group geometric setting 
 is particularly suited to supergravity theories, where local supersymmetry variations can be interpreted as diffeomorphisms in the super Poincar\'e group manifold, along the fermionic directions. It is then clear how to proceed to find theories invariant under local supersymmetry transformations: we must devise a procedure
 that yields actions, invariant under superdiffeomorphisms. This is very similar in spirit to the superspace approach \cite{GGRS,WB}, where  supergravity multiplets of dynamical (and auxiliary) fields are contained into a single superfield, depending on superspace coordinates. However the group manifold approach has important differences, as we explain in the coming Sections.
 
The action is obtained with an algorithmic procedure, as the integral of a $d$-form, ``living" on the whole supergroup (soft) manifold $\Gtilde$, but integrated on a $d$-dimensional bosonic {\it submanifold} of $\Gtilde$. This leads to an ordinary spacetime action containing the dynamical fields (and possibly also the auxiliary fields) of a $d$-dimensional supergravity theory. This algorithm will be discussed in detail and 
applied to obtain the action of $N=1$, $d=4$ supergravity. 

The original references, where this approach was first proposed, 
 are given in \cite{gm11}-\cite{gm14}. Reviews can be found 
 in \cite{gm21}-\cite{gm25}, and \cite{EGH} is a standard reference for the use of differential forms
 in gravity and gauge theories.
 
 The paper is organized as follows. In Section 2 we recall the algebraic basis of $d=4$, $N=1$ supergravity as a theory on the (soft) superPoincar\'e manifold, and the passage to a spacetime action. Section 3 deals with the symmetries of the spacetime action, as inherited from the diffeomorphism invariances of the group manifold action. In Section 4 the variational principle is formulated for the group manifold action, and 
 equations of motion are derived. The building rules for (super)group manifold actions are
 discussed in Section 5, and applied to arrive unambiguously at the group manifold Lagrangian for
 $N=1$, $d=4$ supergravity.  Some conclusions, and a selected list of applications and advantages of the group-geometric approach are discussed in Section 6. Finally, the Appendices contain brief accounts of group manifold geometry, integration on supermanifolds and gamma matrix properties.

 \sect{Supergravity from superPoincaré geometry}
 
 \subsection{Soft superPoincar\'e manifold} 
 
Supergravity in first order vierbein formalism can be recast in a supergroup geometric setting as follows.
Consider $G$ = superPoincar\'e group, and denote the vielbein on the $\Gtilde$ manifold as
$\sigma^A = (V^a,\omega^{ab},\psi^\alpha)$. The index $A=(a,ab,\alpha)$ runs on the translations $P_a$, Lorentz rotations $M_{ab}$ and supersymmetry charges $\Qbar_\alpha$ of the superPoincar\'e Lie algebra:
 \eqa
 & &  [P_a,P_b] =0  \label{PoincarePP} \\
 & & [M_{ab},M_{cd}]= -{1 \over 2} (\eta_{ad} M_{bc} + \eta_{bc} M_{ad} -\eta_{ac} M_{bd} -\eta_{bd} M_{ac}) \label{PoincareMM} \\
 & & [M_{ab},P_c]= -{1 \over 2} ( \eta_{bc} P_{a} - \eta_{ac} P_{b}) \label{PoincareMP} \\
 & & [P_a,\Qbar_\al]=0 \\
 & & [M_{ab},\Qbar_\beta]= -{1 \over 4} \Qbar_\alpha (\gamma_{ab})^\alpha_{~\beta}   \label{sPoincareMQ} \\
 & & \{ \Qbar_\alpha,\Qbar_\beta \} = -i (C\gamma^a)_{\alpha \beta}  P_{a}, \label{sPoincareQQ}
  \ena 
$\eta$ being the flat Minkowski metric, and $C_{\al\be} $ the charge conjugation matrix.  The spinorial generator $\Qbar_\alpha  \equiv Q^\beta
C_{\beta\alpha}$ is a Majorana spinor, i.e. $Q^\beta C_{\beta\alpha} = Q^\dagger_\beta (\gamma_0)^\beta_{~\alpha}$.
Thus the super-Poincar\'e manifold has 10 bosonic directions with coordinates $x^a$, $y^{ab}$, parametrizing  translations and Lorentz rotations, and 4 fermionic directions with Grassmann coordinates $\theta^\alpha$, corresponding to the 4 supercharges $\Qbar_\alpha, \alpha=1,..4$. 

The components of the supervielbein of the $\Gtilde$ =(soft) superPoincar\'e manifold are the vierbein $V^a$, the spin connection $\omega^{ab}$ and the
gravitino $\psi^\alpha$.
 corresponding respectively to the generators $P_a$, $M_{ab}$ and $\Qbar_\alpha$,

 Using the structure constants of the Lie superalgebra, the curvature (\ref{Gcurvature})
becomes :
\eqa
& & R^{a}= dV^{a} - \omega^{a}_{~c} V^{c} - \frac{i }{2} \bar\psi\gamma^{a} \psi \equiv \Dcal V^a  - \frac{i }{2} \bar\psi\gamma^{a} \psi\label{RasuperPoincare}\\
& & R^{ab}=d \omega^{ab} - \omega^{a}_{~c} ~\omega^{cb} \label{RabsuperPoincare}\\
& & \rho= d \psi- \frac{1}{ 4} \omega^{ab} \gamma_{ab} \psi \equiv \Dcal \psi \label{rhosuperPoincare}
\ena
defining respectively the supertorsion, the
Lorentz curvature and the gravitino field strength. $\Dcal$ is the Lorentz covariant exterior derivative.
Wedge products between forms are understood when omitted.

Taking the exterior derivative of these definitions yields the Bianchi identities:
\eqa
& & dR^a -\omega^a_{~b} R^b + R^a_{~b} V^b - i \psibar \gamma^a \rho \equiv  \Dcal R^a + R^a_{~b} V^b - i \psibar \gamma^a \rho= 0 \label{BianchiRasuperPoincare}\\
& & dR^{ab} - \omega^a_{~c} R^{cb} +  \omega^b_{~c} R^{ca} \equiv \Dcal R^{ab}=0 \label{BianchiRabsuperPoincare}\\
& & d\rho - {1 \over 4} \omega^{ab} \gamma_{ab} \rho + {1 \over 4} R^{ab} \gamma_{ab} \psi \equiv \Dcal \rho 
+ {1 \over 4} R^{ab} \gamma_{ab} \psi =0 \label{BianchirhosuperPoincare}
\ena
At this stage all the fields depend on all $\Gtilde$ manifold coordinates, corresponding
to the generators of the Lie superalgebra: thus $V^a=V^a (x,y,\theta)$, $\omega^{ab} = \omega^{ab}(x,y,\theta)$, $\psi^\alpha (x,y,\theta)$, where the coordinates $x^a$, corresponding to the translations $P_a$, describe usual spacetime. Moreover the one-forms $V^a$, $\omega^{ab}, \psi$ live on the whole
$\Gtilde$, and therefore can be expanded as:
 \eqa
  & & V^a = V^a_\mu (x,y,\theta) dx^\mu + V^a_{\mu\nu} (x,y,\theta) dy^{\mu\nu} + V^a_\alpha (x,y,\theta) d\theta^\alpha \\
  & & \omega^{ab} = \omega^{ab}_\mu (x,y,\theta) dx^\mu + \omega^{ab}_{\mu\nu} (x,y,\theta)dy^{\mu\nu} 
  +\omega^{ab}_{\alpha} (x,y,\theta)d\theta^{\alpha}\\
  & & \psi^\al=\psi^\al_\mu (x,y,\theta) dx^\mu + \psi^\al_{\mu\nu} (x,y,\theta) dy^{\mu\nu} +\psi^\al_\beta (x,y,\theta) d\theta^\beta 
  \ena

\subsection{Group manifold action}

The overabundance of field components, and their dependence on $y$ and $\theta$ coordinates
can be tamed by defining an appropriate action principle. To end up with a geometrical theory in
four spacetime dimensions, we first construct a 4-form Lagrangian $L$ made out of the 
$\Gtilde$ vielbein $\sigma^A$ and its curvature $R^A$, according to a few
building rules to be discussed in Section 5. The resulting Lagrangian for superPoincar\'e supergravity is given by:
\eq
L =
R^{ab} V^{c} V^{d} \epsilon_{abcd} + 4
\bar\psi\gamma_{5} \gamma_{a} \rho V^{a} 
\en
We then define an action by integrating this Lagrangian on a 4-dimensional submanifold $M^4$
of the $\Gtilde$ manifold, spanned by the $x$ coordinates. 

Integration on submanifolds $M^d$ of a $d$-form $L$ that lives on a $g$-dimensional bigger space $\Gtilde$ is carried out as follows: we multiply $L$ by the {\it Poincar\'e dual} of $M^d$, a (singular) closed  ({\it g-d})-form $\eta_{M^d}$ that localizes the Lagrangian on the submanifold $M^d$, and integrate the resulting $g$-form on the whole $\Gtilde$. Thus the group manifold action has the general expression
\eq
 S= \int_{\Gtilde} L \wedge  \eta_{M^d}    \label{Gintegral}
\en

The fields of the theory are those contained in $L$, i.e. the $\Gtilde$ vielbein components, and the
embedding functions that define the $M^d$ submanifold of $\Gtilde$, present in $\eta_{M^d}$. We will see
later that the embedding functions do not enter the field equations obtained from the variation of (\ref{Gintegral}). This program makes use of standard integration theory when $\Gtilde$ is a bosonic space, but requires some new ingredients when $\Gtilde$ is a supermanifold, discussed in Appendix B.

In our $d=4$ supergravity example the group manifold action is the integral on the 14-dimensional  $\Gtilde$ = soft superPoincar\'e manifold:
\eq
S=  \int_{\Gtilde} (R^{ab}V^c V^d \epsilon_{abcd}  + 4
\bar\psi\gamma_{5} \gamma_{a} \rho V^{a})~  \eta_{M^4}    \label{GintegralSP}
\en

\subsection{Spacetime action}

A spacetime action, i.e. an action that is the integral on $M^4$ of a Lagrangian containing fields depending only on $x$, is obtained from  (\ref{GintegralSP}) with a {\it particular choice} of  $\eta_{M^4}$. Integration on
$y$ and $\theta$ coordinates produces then the spacetime action. This particular Poincaré dual is the product of two pieces: $\eta_{M^4}= \eta_y  \wedge \eta_\theta$, where $\eta_y$ is a (singular) 10-form that  localizes
the Lagrangian on the $y=0$ hypersurface:
 \eq
  \eta_{y} =  \delta (y^{12}) \delta (y^{13}) \cdots \delta (y^{34}) dy^{12} \wedge dy^{13} \wedge \cdots \wedge dy^{34}  \label{eta6form}
  \en
 Integration on the $y$ coordinates reduces (\ref{GintegralSP}) to an integral on the superspace $M^{4|4}$ spanned by $x$ and $\theta$. The $y$
 dependence of all fields in $L$ disappears because of the delta functions in $\eta_y$, and
 the ``legs" of $L$ along $dy$ differentials are killed by the product of all independent $dy^{\mu\nu}$ in $\eta$. The other piece of $\eta_{M^4}$ (see Appendix B), after integration on $\theta$ coordinates, produces
 an integral on $M^4$ of a Lagrangian $4$-form, not depending any more on the $\theta$ and on the $d\theta$ differentials.
 Thus the action
 \eqa
 & & S_{spacetime}= \int_{\Gtilde} L \wedge \eta_{M^4} = \int_{M^4} L_{y=0,dy=0,\theta=0,d\theta=0} \nonumber \\
 & & ~~~~~~~~~~~~ = \int_{M^4} R^{ab}V^c V^d \epsilon_{abcd}  + 4
\bar\psi\gamma_{5} \gamma_{a} \rho V^{a} \label{M4integral}
 \ena
contains only the usual fields $V^a_\mu(x)$ and $\omega^{ab}_\mu (x)$ and $\psi(x)$ of $N=1$ supergravity, and reproduces the first order supergravity action.
 \sk
  \noi {\bf Note 1:} $\eta_y$ is closed (because it contains ``functions" depending on $y$ multiplied by all the $dy$ differentials) and not exact (because of the Dirac deltas $\delta(y)$), and thus belongs  to a nontrivial de Rahm cohomology class. In general deformations of
the $M^4$ surface generated by diffeomorphisms leave the Poincar\'e dual $\eta$ in the
same cohomology class, since the Lie derivative commutes with the exterior derivative.
  \sk
    \noi {\bf Note 2:} 
    We will always assume that integration on the Lorentz coordinates
has been carried out, so that all fields depend only on $x$ and $\theta$ coordinates. Moreover all curvatures
are taken to be horizontal in the Lorentz directions. As a consequence  the theory lives in a superspace $M^{4|4}$ spanned by four bosonic coordinates $x^a$ and four fermionic coordinates $\theta^\alpha$.
\sk
\noi {\bf Note 3:} the spacetime action (\ref{M4integral}) and its invariance under the supersymmetry transformations (\ref{susy1})-(\ref{susy3}) were first found in ref. \cite{FFvN} in second
order formalism and in \cite{DZ} in first order formalism, see also the standard references 
 \cite{PvNreport,FVP} on supergravity.

  \sect{Symmetries}
  
  The action (\ref{GintegralSP}) is the integral on $\Gtilde$ of a top form: it is clearly invariant
  under diffeomorphisms on $\Gtilde$. But what we are really interested in are the symmetries 
  of the spacetime action as given in (\ref{M4integral}), where the variations are carried out only in the $x$-dependent fields in $L|_{y=dy=0,\theta= d\theta=0}$. The only symmetries guaranteed a priori 
   are the 4-dimensional spacetime diffeomorphisms, the spacetime action being an integral of a 4-form on $M^4$.
  
  Here resides most of the power of the group manifold formalism: if one considers
  the ``mother"  action (\ref{Gintegral}) on $\Gtilde$, the guaranteed symmetries are {\it all} the diff.s on $\Gtilde$, generated by the Lie derivative $\ell_\epsi$ along the tangent vectors  $\epsi = \epsi^A t_A$ of $\Gtilde$.
  But how do these symmetries transfer to the spacetime action ?  
  
  The variation of the group manifold action under diff.s generated by $\ell_\epsilon$ is\footnote{Recall $\ell_\epsi =  \iota_\epsi d + d \iota_\epsi$ so that $\ell_\epsi$(top form) = $d(\iota_\epsi$ top form)}
  \eq
   \delta S = \int_{\Gtilde}  \ell_\epsi (L \wedge \eta )= \int_{\Gtilde}  (\ell_\epsi L) \wedge \eta + L \wedge \ell_\epsi \eta =0
   \en
  modulo boundary terms. One has to vary the fields\footnote{Since $\ell_\epsi$ satisfies the Leibniz rule,   $\ell_\epsi L$ can be computed by varying in turn all fields inside $L$.} in $L$ as well as the submanifold embedded in $\Gtilde$:  the sum of these two variations gives zero\footnote{In the following the vanishing of action variations will always be understood modulo boundary terms.} on the group manifold action $S$.  
    But what we need
   in order to have a {\it spacetime} interpretation of all the symmetries of $S$, is really
   \eq
   \delta S = \int_{\Gtilde} (\ell_\epsi L) \wedge \eta =0 \label{spacetimesymm}
   \en
 If this holds, varying the fields $\phi$ inside $L$ with the Lie derivative $\ell_\epsilon$ as in (\ref{Lieder}), and then projecting on spacetime,  yields spacetime variations
 \eq
\de \phi (x) =  \ell_\epsi \phi (x,y,\theta) |_{x}
 \en
  that leave the spacetime action (\ref{M4integral}) invariant. We have denoted by $|_x$ the projection on spacetime due to the integration on $y$ and $\theta$ coordinates in (\ref{spacetimesymm}). We call these
  variations {\it spacetime invariances}, since they leave invariant the spacetime action.  They originate from the diff. invariance of the group manifold action,
 and give rise to symmetries of the spacetime action  (\ref{M4integral}) only when (\ref{spacetimesymm})
 holds. This happens if one of the following conditions is satisfied:
 \sk
 \noi $\bullet$ the Lie derivative on $\eta$ vanishes:
 \eq
 \ell_\epsi \eta = 0 \label{elloneta}
 \en
 \sk
\noi  $\bullet$ the spacetime projection of the Lie derivative of $L$ is exact:
 \eq
 ( \ell_\epsi L )|_{x} = d \alpha  \label{ellonL}
  \en
  \noi In this case the variation (\ref{spacetimesymm})
  \eq
  \delta S = \int_{\Gtilde}  (\ell_\epsi L) \wedge \eta =\int_{M^4}    (\ell_\epsi L )|_{x}
  \en
  vanishes after integration by parts.
  The requirement  (\ref{ellonL}) is equivalent to 
  \eq
 ( \iota_\epsi dL )|_{x} = d \alpha'  \label{idonL}
  \en
 since $\l_\epsi = \iota_\epsi d + d \iota_\epsi$.  
 \sk
 \noi The Lagrangian $L$ depends on the $\Gtilde$-vielbein $\sigma^A$ and its curvature $R^A$, so that also $dL$, after use of Bianchi identities, is expressed in terms of $\sigma^A$ and $R^A$. Then condition (\ref{idonL})  translates
 into a {\it condition on the contractions} $\iota_\epsi R^A$, i.e. a condition on the curvature components.
 
 Let us see how this works for superPoincar\'e supergravity. 

  \subsection{Symmetries of $d=4$ supergravity}

  The symmetries of the spacetime action  (spacetime invariances) are those
generated by a Lie derivative $\ell_\epsi$ such that $\iota_\epsi dL|_{x}= d\alpha'$,
cf. (\ref{idonL}). We need to compute $dL$. Using the Bianchi identities 
(\ref{BianchiRabsuperPoincare}) and (\ref{BianchirhosuperPoincare}), and the definition 
of the torsion $R^a$ in (\ref{RasuperPoincare}) we find:
\begin{align}
& dL= 2 R^{ab} R^c V^d \epsi_{abcd} + i R^{ab} \psibar \ga^c \psi V^d \epsi_{abcd}+
4 \rhobar \ga_5 \ga_a \rho V^a + \nonumber \\
&  ~~~~~~~ + \psibar \ga_5 \ga_c \ga_{ab} \psi R^{ab} V^c -4 \psibar \ga_5 \ga_a \rho R^a - 2i \psibar \ga_5 \ga_a \rho \psibar \ga^a \psi \label{dL1}
\end{align}
The gamma matrix identity
\eq
\ga_c \ga_{ab} = \eta_{ac} \ga_b - \eta_{bc} \ga_a +i \epsi_{abcd}\ga_5 \ga^d 
\en
implies $\psibar \ga_5 \ga_c \ga_{ab} \psi =i \epsi_{abcd} \psibar \gamma^d \psi$, so that the second and the fourth term cancel in 
(\ref{dL1}). Moreover from the Fierz identity in Appendix C one deduces
\eq
\gamma_a  \psi \psibar \gamma^a \psi =0 \label{fierz1}
\en
and since $\psibar \gamma_5 \gamma_a \rho = \rhobar \gamma_5 \gamma_a \psi$ also the last term in
(\ref{dL1}) vanishes due to (\ref{fierz1}). Therefore
\eq
dL= 2 R^{ab} R^c V^d \epsi_{abcd} + 4 \rhobar \ga_5 \ga_a \rho V^a 
  - 4 \psibar \ga_5 \ga_a \rho R^a 
\en
\sk
\noi {\bf Lorentz gauge transformations}
\sk
It is immediate to see that if all curvatures are horizontal in the Lorentz directions
(no ``legs" along $\om$) then indeed $\iota_{\epsi^{ab} t_{ab}}dL=0$, and Lorentz
transformations are a spacetime invariance of the supergravity action. This is essentially due to the absence of bare $\omega^{ab}$ in $L$. The general
diffeomorphism formula (\ref{Lieder}) yields then the usual Lorentz transformations
 \eqa
 & & \ell_{\epsi^{cd} t_{cd}} V^a =   \epsi^a_{~b} V^b  \label{LorentzonV2}\\
 & & \ell_{\epsi^{cd} t_{cd}}  \omega^{ab} = d \epsi^{ab} - \omega^a_{~c} \epsi^{cb} + \omega^b_{~c} \epsi^{ca}= \Dcal \epsi^{ab} \label{Lorentzonom2} \\
 & &  \ell_{\epsi^{cd} t_{cd}} \psi = {1 \over 4} \epsi^{ab} \gamma_{ab} \psi
 \ena
We can check directly the invariance of the action under these variations: all curvatures and vierbeins appearing in (\ref{M4integral}) transform
homogeneously, and Lorentz indices are contracted with Lorentz invariant tensors.
\sk
 \noi {\bf Spacetime diffeomorphisms}
 \sk
 \noi Ordinary diff.s along tangent vectors $\partial_\mu$ dual to $dx^\mu$ are invariances of the spacetime action, since (\ref{M4integral}) is an integral on a 4-dimensional manifold of a 4-form.
 \sk
  \noi {\bf Supersymmetry transformations}
  \sk
  \noi Diff.s along tangent vectors $t_\alpha$ dual to $\psi^\alpha$ are spacetime invariances
  provided $\iota_\epsilon dL|_{x}= total ~derivative$ with $\epsilon = \epsilon^\alpha t_\alpha$, that is to say
  \eqa
 & &   \iota_\epsilon dL = 2 (\iota_\epsilon R^{ab}) R^c V^d \epsi_{abcd} + 2 R^{ab} (\iota_\epsilon R^c ) V^d \epsi_{abcd} + 8 \rhobar \gamma_5 \gamma_a (\iota_\epsilon \rho) V^a \nonumber \\
 & & - 4 \epsilonbar \gamma_5 \gamma_a \rho R^a - 4 \psibar \gamma_5 \gamma_a (\iota_\epsilon \rho) R^a
 - 4 \psibar \gamma_5 \gamma_a \rho (\iota_\epsilon R^a) = tot.~ der. \label{idL2}
   \ena
once projected on spacetime.  This is a condition for the contractions on the curvatures, and it is satisfied by:
 \eqa
 & & \iota_\epsilon R^a =0 \label{rh1}\\
 & & \iota_\epsilon R^{ab} = - \epsi^{abef} \rhobar_{ef} \ga_5 \ga_g \epsilon V^g - \epsi^{efg[a} \rhobar_{ef} \ga_5 \ga_g \epsilon V^{b]}  \equiv \thetabar^{ab}_c \epsilon V^c \label{rh2}\\
  & & \iota_\epsilon \rho =0 \label{rh3}
 \ena
Thus we have supersymmetry invariance of the spacetime action if the curvatures have the following parametrization on a basis of 2-forms:
 \eqa
 & & R^a=R^a_{~bc} ~V^b V^c  \label{param1}\\
 & & R^{ab}= R^{ab}_{~~cd} V^c V^d + \thetabar^{ab}_c ~\psi ~V^c  \label{param2}\\
 & & \rho= \rho_{ab}~ V^a V^b \label{param3}
 \ena
 where we have taken into account also horizontality in the Lorentz directions. The conditions (\ref{rh1})-(\ref{rh3}) are called ``rheonomic conditions", and similarly (\ref{param1})-(\ref{param3}) are called
 ``rheonomic parametrizations" of the curvatures.
 
  The diff.s along $\epsilon=\epsilon^\alpha t_\alpha$ (supersymmetry transformations) act on the fields according to the general formula
   (\ref{Lieder}), where the contractions on the curvatures are given in (\ref{rh1})-(\ref{rh3}):
 \eqa
  & & \ell_\epsilon V^a = i \epsilonbar \gamma^a \psi \label{susy1}\\
 & & \ell_\epsilon \om^{ab} =  \thetabar^{ab}_c \epsilon V^c  \label{susy2}\\
 & & \ell_\epsilon \psi = \Dcal \epsilon \equiv d \epsilon - {1 \over 4} \om^{ab} \ga_{ab} \epsilon \label{susy3}
 \ena
with $\thetabar^{ab}_c$ defined in (\ref{rh2}).

 \sect{Variational principle and field equations}
 
The group manifold action (\ref{Gintegral}) is a functional of $L$ and of the embedded submanifold $M$, and
therefore varying the action means varying both $L$ and $M$. Varying $M$
corresponds to varying $\eta_{M}$. Then the variational principle reads:
\begin{equation}
\label{Svariation}{\ \delta S[L, M] =
\int_{\Gtilde} ( \delta L\wedge\eta_{M} + L\wedge\delta\eta_{M}) } =0\,.
\end{equation}
Any (continuous) variation of $M$ can be obtained by acting on 
$\eta_M$ with a diffeomorphism generated by a Lie derivative $\ell_\xi$.
An arbitrary variation is generated by an arbitrary $\xi$ vector, and
the variational principle becomes
\begin{equation}
\label{Svariation1}{\ \delta S[L, M] =
\int_{\Gtilde} ( \delta L\wedge\eta_{M} + L\wedge\ell_\xi \eta_{M}) } =0\,.
\end{equation}
Since field
variations in $L$ and variation of $M$ are independent, the two terms in (\ref{Svariation1})
must vanish separately. From the vanishing of the first one we deduce
\eq
\int_{\Gtilde} (  \delta \phi \we {\part L \over \part \phi}  +  d \delta \phi \we {\part L \over \part (d\phi) } ) \wedge \eta_M =0
\en
where $L=L(\phi,d\phi)$ is considered a function of the 1-form fields $\phi$ and their ``velocities" $d\phi$.  A summation on all fields is understood.
Integrating by parts and recalling $d \eta_M=0$ yields
 \eq
 \int_{\Gtilde} \delta \phi \we ( {\part L \over \part \phi}  + d {\part L \over \part (d\phi) } ) \wedge \eta_M =0
\en
and since the $\delta \phi$ are arbitrary we find 
 \eq
  ({\part L \over \part \phi} + d {\part L \over \part (d\phi)}) \we  \eta_M=0 \label{ELequations0}
  \en
This must hold for any $\eta_M$ (i.e. for generic embedding functions): we arrive therefore at equations
that hold on the whole $\Gtilde$, and are the form version of
  the Euler-Lagrange equations:
  \eq
{\part L \over \part \phi} + d {\part L \over \part (d\phi)}  =0 \label{ELequations1}
\en
If $L$ is a $d$-form, these equations are $(d-1)$-forms. Their content can be examined by
expanding them along a complete basis of  $(d-1)$-forms in $\Gtilde$.

Requiring the vanishing of the second term in the variation (\ref{Svariation1}) does
not imply further equations besides the Euler-Lagrange field equations (\ref{ELequations1}): indeed this term vanishes on the shell of solutions of Euler-Lagrange equations. To prove it, notice that
\eq
\int_{\Gtilde} L \we \ell_\xi \eta_M = - \int_{\Gtilde} \ell_\xi L \we \eta_M =0 ~(on~shell) \label{onshell}
\en
because $\ell_\xi L$ is just a particular variation of $L$, under which the action remains stationary on-shell.

Thus the group manifold variational principle leads to the field equations (\ref{ELequations1}),
holding as $(d-1)$-form equations on the whole $\Gtilde$. 
\sk
\noi {\bf Note 1:} The variational principle 
{\it does not determine} the embedding of $M$ into $\Gtilde$.
\sk
\noi {\bf Note 2:} the field equations (\ref{ELequations1}) are form equations, and therefore
invariant under the action of a Lie derivative. More precisely, if $\phi$ is a solution of
(\ref{ELequations1}), so is $\phi + \ell_\epsi \phi$: Lie derivatives generate symmetries
of the field equations.

\sk
\noi Finally, we have the following 
\sk
\noi {\bf Theorem:} $dL = 0 ~(on~shell) $
\sk
\noi i.e. the Lagrangian, as a $d$-form on $\Gtilde$, is closed on shell. To prove it
recall that
$\eta_M$ is closed , so that  on shell we find, cf. (\ref{onshell}): 
\eq
0 = \int_{\Gtilde} L \we \ell_\xi \eta_M =  \int_{\Gtilde} L \we d \iota_\xi \eta_M =- (-)^d  \int_{\Gtilde} dL \we \iota_\xi \eta_M
\en
$\xi$ being arbitrary, this implies $dL=0$ (on shell)\footnote{
In fact, this is just Stokes theorem applied to a region of $\Gtilde$ bounded by two different hypersurfaces $M$ and $M'$.} \square

Let us apply the preceding discussion to the superPoincar\'e supergravity example.

\subsection{Supergravity field equations}

The variational equations (\ref{ELequations1}) for the group manifold action (\ref{GintegralSP}) read:
\eqa
& & 2 R^{c}  V^{d} \epsilon_{abcd}  = 0   \label{sPoincarefieldeqRa}\\
& & 2 R^{ab}  V^{c} \epsilon_{abcd} + 4 \psibar \ga_5 \ga_d \rho = 0 \label{sPoincarefieldeqRab} \\
& & 8 \ga_5 \ga_a \rho V^a -4 \ga_5 \ga_a \psi R^a=0 \label{sPoincarefieldeqrho}
\ena
obtained varying $\om^{ab}, V^d$ and $\psi$ respectively.
The analysis proceeds as follows: we first expand the curvatures on a basis of 2-forms\footnote{assuming
horizontality in the Lorentz directions. This amounts to consider configurations
satisfying the Lorentz horizontality constraints on the curvatures.}
\eqa
& & R^a = R^a_{~bc} V^b V^c + \thetabar^a_{~c} \psi V^c + \psibar K^a \psi \\
& & R^{ab} = R^{ab}_{~~cd} V^c V^d + \thetabar^{ab}_{~~c} \psi V^c + \psibar K^{ab} \psi \\
& & \rho= \rho_{ab} V^a V^b + H_c \psi V^c + \Omega_{\al\be} \psi^\al \psi^\be
\ena
and then insert them into the field equations (\ref{sPoincarefieldeqRa})-(\ref{sPoincarefieldeqrho}). These, being 3-form equations, can be expanded on the basis $\psi\psi\psi$,  $\psi\psi V$, $\psi VV$, $VVV$. Their content is given below (the three lines correspond to the three eq.s of motion):
\sk
\noi$\psi\psi\psi$ sector:
\eqa
& &\Om_{\al\be} =0\\
& & 0=0\\
& & K^a=0
\ena
$\psi\psi V$ sector:
\eqa
& & 2 \psibar K^{ab} \psi V^c \epsi_{abcd} + 4 \psibar \ga_5 \ga_d H_c \psi V^c=0 \label{psipsiV1}\\
& &  ~~~~~~~~~~~~~~~~~~0=0\\
& & ~~~~~~~~~~~~~~~~~~\thetabar^a_{~c}=0
\ena
$\psi V V$ sector:
\eqa
& & 2 \thetabar^{ab}_{~~e} \psi V^e V^c \epsi_{abcd} + 4 \psibar \ga_5 \ga_d \rho_{ab} V^a V^b =0 \label{psiVV1}\\
& &  ~~~~~~~~~~~~~~~~~~0=0\\
& & \ga_5 \ga_a H_b \psi V^b V^a - 4 \ga_5 \ga_c \psi R^c_{~ab} V^a V^b =0 \label{psiVV3}
\ena
$V V V$ sector:
\eqa
& & R^a_{~bc}=0 \\
& & R^{ac} _{~~bc} - {1\over 2} \delta^a_b ~R^{cd}_{~~cd} =0 \\
& & \gamma^a \rho_{ab}=0
\ena
Inserting $R^a_{~bc}=0$ into (\ref{psiVV3}) yields $H_c=0$, which used in (\ref{psipsiV1})
gives $K^{ab}=0$. Thus the only nontrivial relation in the ``outer" projections is (\ref{psiVV1}),
that determines $\theta^{ab}_{~~c}$ to be
\eq
\theta^{ab}_{~~c}=-\epsi^{abef} \rhobar_{ef} \ga_5\ga_c - \delta^{[a}_c \epsi^{b]efg} \rhobar_{ef} \ga_5\ga_g  \label{thetabar1}
\en
in agreement with the $\theta^{ab}_{~~c}$ obtained from the condition (\ref{rh2}). Thus we arrive at the same curvature parametrizations (\ref{param1})-(\ref{param3})
obtained in Sect. 3.1 by requiring spacetime supersymmetry invariance.

Finally, the $VVV$ sector reproduces the (super)torsion equation, and the propagation equations
for the vierbein and the gravitino.

\sk
\noi{\bf Note:} from the torsion equation 
\eq
2 R^a_{\mu\nu} \equiv \part_\mu V^a_\nu - \part_\nu V^a_\mu - \om^a_{~b,\mu} V^b_\nu +
\om^a_{~b,\nu} V^b_\mu - i \psibar_\mu \ga^a \psi_\nu =0
\en
we can express the spin connection in terms of $V$ and $\psi$, recovering second order formalism:
\eqa
& & \om_{ab,\mu} = {1 \over 2} V^\nu_a V^\rho_b \eta_{cd} ~(\part_{[\mu} V_{\nu]}^c V_\rho^d - \part_{[\mu} V_{\rho]}^c V_\nu^d + \part_{[\nu} V_{\rho]}^c V_\mu^d) + \nonumber \\
& &  ~~~~~~~~ + {i \over 4} V_a^\nu V_b^\rho (\psibar_\mu \ga_\nu \psi_\rho+
\psibar_\nu \ga_\rho \psi_\mu-\psibar_\rho \ga_\mu \psi_\nu-(\nu \leftrightarrow \rho)) \label{omsecondorder}
\ena

 \sect{Building rules}
 
 The group geometric approach provides a systematic set of building rules \cite{gm21} for
 constructing Lagrangians of supersymmetric theories:
 \sk
 1) Choose a Lie (super)algebra $G$, containing generators $P_a$ that can be
 associated to $d$ spacetime directions, and a Lorentz-like subalgebra $H$.
Examples are the superPoincar\'e algebras in $d$ dimensions or their uncontracted versions
 (orthosymplectic superalgebras $OSp(N|2^{[d/2]})$). The fields of the theory
 are the vielbein components of the soft group manifold $\Gtilde$.
 \sk
 2) Construct the most general $d$-form on $\Gtilde$, by multiplying (with exterior products) 1-form  vielbein components $\sigma^A$ and 2-form curvatures $R^A$, without bare Lorentz connection and contracting indices with $H$-invariant tensors, so that the resulting Lagrangian 
 is a Lorentz scalar.  
 \sk
 3) Require that the variational equations admit the ``vacuum solution"  $R^A=0$, described by the
 vielbein of the rigid group manifold $G$.
 \sk
4) The construction is greatly helped by scaling properties of the fields, dictated by the structure
 of the Lie (super)algebra $G$, or equivalently by the Cartan-Maurer equations for the $G$ vielbein. Consider for example the superPoincar\'e algebra: it is invariant under the rescalings $P_a \rightarrow \lambda P_a, M_{ab} \rightarrow M_{ab}, \Qbar_\alpha \rightarrow \lambda^{1\over 2} \Qbar_\alpha$. Then the curvature definitions (\ref{RasuperPoincare})-(\ref{rhosuperPoincare}) are invariant under 
 \eq
 V^a \rightarrow \lambda V^a, ~~~\om^{ab} \rightarrow \om^{ab}, ~~~\psi \rightarrow \lambda^{1\over 2} \psi \label{rescalings1}
 \en
 The field equations must be invariant under these rescalings, and therefore the action must scale homogeneously under (\ref{rescalings1}). Since the Einstein-Hilbert term scales as $\lambda^2$, all terms must scale in the same way, and this restricts the candidate terms in the Lagrangian. 
 \sk
 5) Finally, requiring
 that all terms have the same parity as the Einstein-Hilbert term further narrows the list of candidates.
 \sk
 
  \subsection{The Lagrangian for $d=4$ supergravity}
 
 Following the above rules, one arrives at the  $d=4$ supergravity action (\ref{GintegralSP}). We recall here
 the steps of the procedure \cite{gm21}.  The most general lagrangian 4-form satisfying Rule 1 can at most contain two curvatures, and is therefore of the type:
 \eq
 L = R^A R^B \nu_{AB} + R^A \nu_A + \Lambda
 \en
 with\footnote{repeated indices are contracted with the Minkowski flat metric.}
 \eq
 R^A R^B \nu^{(2)}_{AB}=c_1 R^{ab} R^{cd} \epsi_{abcd}+c_2 R^{ab}R^{ab}
+c_3R^a R^a +c_4 \rhob \rho+c_5 \rhob \ga_5 \rho  \label{quadratic}
\en

\noi The first two are total derivatives, and are related to the Euler characteristic and to the 
Pontriagyn number of $M^4$. The last three can be 
reduced to linear terms in the curvatures plus total derivatives. 
Actually scaling invariance  eliminates all the terms in 
(\ref{quadratic}) except 
$R^a R^a$, since the Einstein term scales as $\lambda^2$.
The torsion-squared term can be reduced to a linear term since

\eq
R^a R^a = ({\cal D} V^a - {i \over 2} \psib \ga^a \psi)R^a=d(V^a R^a)+
V^a(-R^{ab}V^b+i\psib\ga^a\rho)-{i \over 2} \psib \ga^a \psi R^a 
\en

\noi in virtue of the Bianchi identity (\ref{BianchiRasuperPoincare}). This leaves us with a 
lagrangian of the form:

\eq
L=\Lambda+\nu_{ab} R^{ab} + \nu_a R^a + {\bar\nu} \rho 
\en

\noi where

\eqa
 & &\Lambda=\alpha_1 \epsi_{abcd} V^aV^bV^cV^d+i\alpha_2\epsi_{abcd}\psib 
\ga^{ab} \psi V^cV^d + i\alpha_3 \psib \ga^{ab} \psi V^aV^b \\
& &\nu_{ab}= \beta_1 \epsi_{abcd} V^c V^d + \beta_2 V^a V^b + i \beta_3 
\psib \ga_{ab} \psi + i \beta_4 \epsi_{abcd} \psib \ga^{cd} \psi \\
& &\nu_a=i \eta_1 \psib \ga_a \psi\\
& &\nu=\de_1 \ga_5 \ga_a \psi V^a + i \de_2 \ga_a \psi V^a 
\ena
 
\noi are the most general Lorentz covariant terms. Notice that the only 
nonvanishing $\psi\psi$ currents are $\psib \ga^a \psi$ and $\psib \ga^
{ab} \psi$. Correct $\lambda^2$ 
scaling of $L $ drastically reduces the possible 
terms:
$\alpha_1=\alpha_2=\alpha_3=\beta_3=\beta_4=0$. Moreover parity 
 implies $\beta_2=\eta_1= \de_2=0$ (all terms must 
have the same parity as the Einstein term $R^{ab} V^c V^d \epsi_{abcd}$, 
i.e. must be pseudoscalars). Thus we finally have:

\eq
L=\beta_1 \epsi_{abcd} R^{ab} V^c V^d + \de_1 \psib \ga_5 \ga_a 
\rho V^a  \label{Lagrangian2}
\en

\noi The requirement 3) that 
the vacuum be a solution of the field equations fixes the last parameter
$a=\de_1/ \beta_1$. Indeed the field equations obtained from the Lagrangian
(\ref{Lagrangian2}) by varying $V^a, \om^{ab}$ and $\psi$ are respectively:

\eqa
& & 2 R^{ab} V^c \epsi_{abcd} + a \psib \ga_5 \ga_d \rho = 0 \\
& & 2{\cal D} V^c V^d \epsi_{abcd} + {1 \over 4} a \psib \ga_5 \ga_d 
\ga_{ab} \psi V^d =0  \label{fieldeq2} \\
& & 2a\ga_5\ga_a \rho V^a - a \ga_5 \ga_a \psi R^a = 0
\ena

\noi To find the first is immediate; for the second 
we only have to recall 
that varying $\om^{ab}$ in $R^
{ab}$ yields $\de  R^{ab}={\cal D}(\de \om^{ab})$, and that by 
integrating by parts the Lorentz covariant derivative ${\cal D}$ 
can be transferred on $V^a$. Finally for the gravitino variation we have

\eqa
& & {1 \over a} \de {\cal L}
= (\de \psib) \ga_5 \ga_a {\cal D} \psi V^a +  \psib \ga_5 
\ga_a {\cal D} (\de \psi) V^a =\\
& & =(\de \psib) \ga_5 \ga_a {\cal D} \psi V^a+ \psib \ga_5 \ga_a \de\psi 
{\cal D} V^a+ \de\psib \ga_5 \ga_a {\cal D} \psi V^a=\\
& & = 2 (\de \psib) \ga_5 \ga_a {\cal D} \psi V^a - \de \psib \ga_5 \ga_a \psi 
( R^a + {i \over 2} \psib \ga^a \psi ) =\\
& &=(\de \psib) (2 \ga_5 \ga_a {\cal D} \psi V^a-\ga_5 \ga_a \psi R^a)\\
\ena

\noi in virtue of $\psib \ga_5 \ga_a (\de\psi)=-(\de \psib) 
\ga_5 \ga_a \psi$ 
and the Fierz identity

\eq
\ga_a \psi \psib \ga^a \psi=0 
\en

\noi Note that using the gamma-algebra identity:

\eq
\ga_5 \ga_d \ga_{ab}=2\ga_5 \de_{d[a} \ga_{b]}-i \epsi_{abcd} \ga^c 
\en

\noi the variational equation (\ref{fieldeq2}) can be recast in the form:

\eq
2R^cV^d\epsi_{abcd}+{i \over 4} (4-a) \psib \ga_5 \ga_d 
\ga_{ab} \psi V^d =0 
\en
\noi so that the vacuum, defined by vanishing curvatures, is a solution 
of the field equations (5.21) only if $a=4$.
\sk
In conclusion: applying  the building rules with $G$ = superPoincaré yields  the $N=1$, $d=4$ supergravity action (\ref{GintegralSP}).

\sect{Conclusions}

In this review we have focused mostly on the logic of the group manifold approach, applied to a single example,
i.e. $N=1$, $d=4$ supergravity. Comprehensive discussions on the applications of the method
for the construction of supergravity theories in diverse dimensions can be found in the recent reviews \cite{gm24,gm25}. 
\sk
\noi We list here some of the advantages/motivations:
\sk
\noi - all fields have a group-geometric origin, even if they are not all
 gauge fields.
 
 \noi - all symmetries have a common origin as diffeomorphisms on $\Gtilde$.
 
\noi  - there is a systematic procedure based on group geometry to construct
  actions, invariant under diffeomorphisms, and under gauge symmetries closing
  on a subgroup of $G$.
  
\noi   - supersymmetry is formulated in a very natural way as a
  diffeomorphism in Grassmann directions of a supermanifold.
  
\noi - closer contact is maintained with the usual component actions, whereas in the superfield formalism the actions looks quite different.
  In fact the group manifold action interpolates between the component and the superfield actions of the same supergravity theory, see \cite{if1}-\cite{if3}.

\noi - in the group manifold formulation of $d=6$ supergravity \cite{d6SG} and $d=10$ supergravity
\cite{d10SG} the selfdual conditions for the 3-form (in $d=6$) and 5-form (in $d=10$) curvatures are a yield of the field equations in the respective superspaces, and do not need to be imposed as external constraints.
\sk
\noi Finally, we recall some conceptual advances due to the group-geometric treatment of supergravity:
\sk
\noi - the generalization to $p$-form
potentials, necessary to treat supergravity theories with $p$-form fields, in the framework of 
Free Differential Algebras (FDA) \cite{sullivan,DFd11,gm21,DFTvN,FDAnew1}, and their dual formulation \cite{FDAdual1} - \cite{FDAdual4}. 

\noi - the bridge between superspace and group manifold methods provided by 
superintegration, developed in ref.s \cite{if1}-\cite{if3}.

\noi - a covariant hamiltonian formalism, initially proposed in \cite{CCF1}-\cite{CCF3}, based on the definition of field momenta as
derivatives of the Lagrangian with respect to the exterior derivative of the fields, not involving
a preferred direction (time). Recent developments \cite{CD,SGcovariantH} include the construction
of all canonical symmetry generators for $N=1$, $d=4$ supergravity \cite{SGcovariantH}. This
covariant hamiltonian formalism can also be generalized to a noncommutative (twisted) 
setting \cite{LCtwistedH}, describing noncommutative twisted (super)gravity \cite{AC1,AC2}.

\section*{Acknowledgements}

In writing this review we have benefited from discussions with Carlo Alberto Cremonini, Riccardo D' Auria and Pietro Antonio Grassi. We acknowledge partial support from INFN, CSN4, Iniziativa Specifica GSS.  This research has a financial support from Universit\`a del Piemonte Orientale.

\appendix

 \sect{Group manifold geometry}

This brief resum\'e is taken from Sec. 2 of \cite{gm23}.
We start from a Lie algebra Lie({\sl G}), with generators $T_A$ 
satisfying the commutation relations
\eq
[T_A,T_B]={C^C}_{AB}T_C  \label{2.1}
\en
\noi For simplicity we consider only usual Lie algebras. The extension to superalgebras
is straightforward and only necessitates extra signs (for ex. anticommutators
for fermionic generators) due to gradings. 

A generic group element $g \in G$ connected with the identity
\footnote{Hereafter $G$ indicates the part of the group connected 
with the identity.} can be expressed as
\eq
g=\exp (y^A T_A) \equiv y \label{2.2}
\en
\noi where $y^A$ are the (exponential) coordinates of the group 
manifold. Each element of $G$ is labelled by the coordinates $y^A$, and 
for notational economy we denote it simply by $y$. Similarly $yx$ stands 
for ~$\exp (y^A T_A) \exp (x^B T_B)$, the product of two group elements, 
and by $(yx)^M$ we denote the corresponding coordinates.

Consider now $(yx)^M$ as a function\footnote{Since $G$ is a Lie 
group, this function is smooth.} of $x^A$:
\eq
(yx)^M=y^M+e_A^{~~M} (y) x^A + e_{AB}^{~~~M} (y) x^A x^B+ ... \label{2.3}
\en
\noi For infinitesimal $x$:
\eq
(yx)^M=y^M+(x^A t_A) y^M= (1+x^A t_A) y^M, ~~~t_A 
\equiv e_A^{~~N} (y) \pdyN \label{2.4}
\en
\noi so that the $t_A$ are a differential representation of the 
abstract generators $T_A$, and satisfy therefore the same algebra:
\eq
[t_A, t_B]=\CCAB t_C \label{2.5}
\en
The geometrical meaning of the components $e_A^{~N} (y)$ in eq. (\ref{2.3}) 
is clear: consider the infinitesimal displacement $\de_A y^M$ due to the 
(right) action of $1+ \varepsilon T_A$ ($\epsi$ = infinitesimal 
parameter). Then 
\eq
\de_A y^M = \epsi e_A^{~~M} (y) \label{2.6}
\en
\noi and the dim$G$ vectors $e_A^{~~M} (y)$, A=1,...dim$G$ are simply 
the tangent vectors at $y$ in the direction of the displacements $\de_A 
y^M$. It is customary to call tangent vector along the $T_A$ 
direction the whole differential operator $t_A \equiv 
e_A^{~~N} (y) \pdyN$. 

Note that $e_A^{~~M}$ is an invertible matrix, 
since the map $y \rightarrow yx$ is a diffeomorphism.

The $t_A (y)$ span the tangent space of $G$ at $y$: they form a 
contravariant basis. The ``coordinate" 
basis given by the vectors $\pdyN$
is related to the $t_A$ (the intrinsic basis) 
via the nondegenerate matrix $e_A^{~~N}$.
The indices A,B,... are tangent space indices (``flat" indices) and are
inert under $y$ coordinate transformations. The indices M,N,... are
coordinate indices (``world" indices) and do transform under coordinate
transformations in the usual way (see later).   
Next we define the one-forms $\si^A (y)$ as the duals of the $t_A$:
\eq
\si^A (t_B) = \de_A^B \label{2.8}
\en
\noi The $\si^A$ are a covariant basis (the intrinsic vielbein basis)
for the dual of the tangent space, called 
cotangent space (the space of 1-forms). The ``coordinate" 
cotangent basis dual to the $\pdyN$ vectors is given by the 
differentials $dy^M$ ($dy^M(\pdyN)=\de^M_N$). The components of 
$\si^A(y)$ on the coordinate basis are denoted $e_M^{~~A} (y)$:
\eq
\si^A (y)=e_M^{~~A} (y)~ dy^M   \label{2.9}
\en
\noi From the duality of the tangent and cotangent bases:
\eqa
& & e_M^{~~A}~e_B^{~~M}=\de_B^A  \\
& & e_A^{~~M}~e_N^{~~A}=\de_N^M \label{2.10}
\ena
\sk
\noi {\bf Note 1}: Substituting $t_A$ by
$e_A^{~~N} (y) \pdyN$  into the commutator (\ref{2.5}) leads to
the differential condition on $e_A^{~~M} (y)$:
\eq
 -2e_{[A}^{~~~N}~e_{B]}^{~~~M} \partial_N e_M^{~~C} = \CCAB
 \label{2.11}
 \en
{\bf Note 2}: computing the exterior derivative of $\si^A$, using eq.s 
(\ref{2.9}) and (\ref{2.11}) leads to the equations
\eq
d\si^A+{1 \over 2}{C^A}_{BC}\si^B \we \si^C = 0 \label{2.12}
\en
\noi These are called {\sl Cartan-Maurer equations}, and provide a dual 
formulation of Lie algebras in terms of the one-forms $\si^A$. It is 
immediate to verify that the closure of the exterior derivative ($d^
2=0$) is equivalent to the Jacobi identities for the structure 
constants:
\eq
C^A_{~~B[C} C^B_{~~DE]} =0  \label{2.13}
\en
\noi (apply $d$ to eq. (\ref{2.12})).
\sk
\noi {\bf Note 3}:
\sk
\noi Defining $\si (y) \equiv \si^A (y) T_A$ 
the Cartan-Maurer eq.s (\ref{2.12}) 
take the form
\eq
d\si+\si\we\si=0 \label{2.14}
\en
The Lie-valued one-form $\si(y)$ can also be constructed directly 
from the group element $y$:
\eq
\si(y)=y^{-1} dy  \label{2.15}
\en
It is easy to verify that (\ref{2.15}) satisfies the Cartan-Maurer equation (\ref{2.14}) (use $dy^{-1}=-y^{-1}dy~y^{-1}$). Moreover, it takes the same value 
as $e_M^{~~A} dy^M~T_A$ at the origin $y=0$. Indeed from the definition 
of $e_A^{~~M}$ in eq. (\ref{2.3}) one sees that $e_A^{~~M} (y=0)= \de_A^M$, 
and therefore $e_M^{~~A} (0) dy^M ~T_A=dy^A~T_A$. This value coincides 
with $y^{-1}dy|_{y=0}$ since $y^{-1}|_{y=0} =$[group unit], and 
$dy|_{y=0}=dy^A T_A$ (from (\ref{2.2})). This 
observation suffices to conclude that $y^{-1} dy$ is equal to 
$e_M^{~~A} (y) dy^M T_A$.
\sk
\noi {\bf Soft group manifold}
\sk
Consider a smooth deformation $\Gt$ of the group manifold $G$. Its 
vielbein field is given by the intrinsic cotangent basis, 
defined for any differentiable manifold:
\eq
\mu^A (y)=\mu_M^{~~A} (y) dy^M \label{3.1}
\en
\noi (In this Appendix we use the symbol $\mu$ for the 
``soft" vielbein). In general $\mu^A$ does not satisfy the Cartan-Maurer 
equations any more, so that
\eq
d\mu^A+{1 \over 2} \CABC \mu^B \we \mu^C \equiv R^A \not= 0 \label{3.2}
\en
\noi The extent of the deformation $G \rightarrow \Gt$ is measured by 
the curvature two-form $R^A$. $R^A = 0$ implies $\mu^A=\si^A$ and 
viceversa.
\sk
Applying the external derivative $d$ to the definition (\ref{3.2}), using $d^2
=0$ and the Jacobi identities on $\CABC$, yields the Bianchi identities
\eq
(\nabla R)^A \equiv dR^A - \CABC R^B \we \mu^C =0 \label{3.3}
\en
\sk
\noi {\bf Diffeomorphisms and Lie derivative}
\sk
First we discuss the variation under diffeomorphisms of the vielbein
field $\mu^A(y)$:
\eqa
     & & \mu^A (y+\de y)-\mu^A(y)=\de [\mu_M^{~~A}(y)dy^M]= \nonumber \\
     & & =(\part_N \mu_M^{~~A}) \de y^N dy^M+ \mu_M^{~~A} (\part_N \de y^M) 
        dy^N= \nonumber\\
     & &=dy^N[\part_N \de y^A+\de y^M(\part_M \mu_N^{~~A}-\part_N \mu_M
       ^{~~A})]= \nonumber\\
     & & = d \de y^A-2 \mu^B \de y^C (d\mu^A)_{BC}=d(\iota_{\de y}\mu^A)
       +\iota_{\de y} d \mu^A \label{3.7} 
       \ena
\noi where
\eq
\de y^A \equiv \de y^M \mu_M^{~~A},~~~\de y \equiv \de y^M 
\part_M,~~~d\mu^A\equiv (d\mu^A)_{BC} \mu^B \we \mu^C, \label{3.8}
\en
\noi and the contraction $\iota_{t}$ along a tangent vector ${t}$ 
is defined on p-forms
\eq
\om_{(p)}=\om_{B_1...B_p}\mu^{B_1} \we ... \we \mu^{B_p} 
\en
\noi as  
\eq
\iota_{t}~ \om_{(p)}=p~ t^A \om_{AB_2...B_p}\mu^{B_2}
 \we ... \we \mu^{B_p}
    \label{3.9}
    \en
\noi Note that $\iota_{t}$ maps p-forms into ($p-1$)-forms. 
The operator 
\eq
\ell_{t} \equiv d~ \iota_{t} + \iota_{t} ~d  \label{3.10}
\en
\noi is called the {\sl Lie derivative} along the tangent vector $t$ 
and maps p-forms into p-forms. 
As shown in eq. (\ref{3.7}), the Lie derivative of the one-form 
$\mu^A$ along $\de y$ gives 
its variation under the diffeomorphism $y \rightarrow y+\de y$.   
This holds true for any p-form.
\sk
We now rewrite the variation $\de \mu^A$ of eq. (\ref{3.7}) in a suggestive 
way, by adding and subtracting $\CABC \mu^B  \de y^C$ :
\eqa
   \de \mu^A &= & d \de y^A + \CABC \mu^B  \de y^C - 2 \mu^B \de y^C 
(d\mu^A)_{BC} - \CABC \mu^B \de y^C \\
             &=& (\nabla \de y)^A + \iota_{\de y} R^A \\
      \label{3.11}
      \ena
\noi where we have used the definition (\ref{3.2}) for the curvature, and
the $G$-covariant derivative $\nabla$ acts on $\de y^A$ as
\eq
(\nabla \de y)^A \equiv d \mu^A + \CABC \mu^B \de y^A \label{3.12}
\en

{\bf The algebra of Lie derivatives}
\sk
The algebra of diffeomorphisms
 is given by the commutators of Lie derivatives:
 \eq
 \left[ \ell_{ \epsi^A_1 t_A},\ell_{\epsi_2^B t_B} \right]= \ell_{\epsi^C_3 t_C} \label{aldiffLie}
\en
with
\eq
\epsi_3^C=\epsi^A_1 \partial_A
\epsi^C_2 - \epsi^A_2 \partial_A \epsi^C_1 - 2 \epsi^A_1 \epsi^B_2
\Rcal^C_{AB}
\en
and
 \eq \Rcal^C_{AB} \equiv R^C_{AB}-\unmezzo C^C_{AB}
\label{Rcal}
\en
The components $R^A_{BC}$ are defined by $R^A = R^A_{BC} \mu^B \we \mu^C$. The closure of the algebra requires the Bianchi identities (\ref{3.3}), that we can rewrite in the form
\eq
\part_{[B} \Rcal^A_{CD]}+2~\Rcal^A_{E[B} \Rcal^E_{CD]} =0
\label{BianchiRcal}
\en
To prove (\ref{aldiffLie}) just apply both sides of the equation to the basic (soft) vielbein $\mu$.
\sk
\sk

\sect{Integration on supermanifolds: integral forms}

In this Appendix, taken from Section 4 of \cite{gm24},  we recall basic results in supermanifold integration
 (see for ex. \cite{w1} for a recent review, or \cite{voronov} for a textbook),
 and new developments concerning integral forms, discussed in ref.s  \cite{if1}-\cite{if3}.
 
We have defined the supergravity action (\ref{GintegralSP})  as an integral of a top form on the superPoincar\'e group manifold. We have given explicitly only the 4-form Lagrangian, postponing 
the precise expression of $\eta_{M^4}$ to the present Section. In fact in the supergravity case we have tacitly assumed typical properties of bosonic integration, as for ex. the existence of a top form and Stokes' theorem. Here we want to justify these assumptions, and give a short
account of superintegration theory.

The construction of actions invariant under diffeomorphisms is solved 
``ab initio" in ordinary integration theory by {\it form} integration. The integral of a $d$-form
\eq
\omega^{(d)} =\omega_{[\mu_1 \cdots \mu_d]} (x) ~dx^{\mu_1} \wedge \cdots \wedge dx^{\mu_d} \label{dform}
\en
on a $d$-dimensional manifold $M^d$ is defined by
\eq
I =   \int_{M^d} \om^{(d)}  \equiv \int_{M^d} {1 \over d!}~ \omega_{[\mu_1 \cdots \mu_d]} (x) \label{intform}\epsilon^{\mu_1 \cdots \mu_d}  ~d^dx
\en
i.e. by usual (Riemann-Lebesgue) integration on $M^d$ of the function $ {1 \over d!}~ \omega_{[\mu_1 \cdots \mu_d]} (x)\epsilon^{\mu_1 \cdots \mu_d} $, where 
$\epsilon^{\mu_1 \cdots \mu_d}$ is the Levi-Civita antisymmetric symbol in the coordinate basis,
a tensor density of weight $-1$. Therefore
\eq
 \epsilon^{\mu_1 \cdots \mu_d} ~d^dx = \epsilon^{\mu_1 \cdots \mu_d} ~dx^1 \we \cdots \we dx^d
 \en
 is a tensor, and the integrand of (\ref{intform}) is a scalar.
 
 As in the previous Sections, we can consider infinitesimal diffeomorphisms as
 active transformations, generated by the Lie derivative $\ell_\epsilon = \iota_\epsilon d + d \iota_\epsilon $.
 Then the form integral (\ref{intform}) transforms as
 \eq
 \delta I = \int_{M^d} \ell_\epsilon \om^{(d)} =  \int_{M^d} (\iota_\epsilon d + d \iota_\epsilon ) \om^{(d)} =0
 \en
 since $d  \om^{(d)} =0$ ($ \om^{(d)}$ is a top form) and $\int_{M^d} d (\iota_\epsilon \om) =0$ for appropriate boundary conditions. Thus we have checked invariance of the form integral under infinitesimal diff.s
 generated by the Lie derivative. Note that the existence of a top form, namely the fact that a $d$-form is closed on $M^d$, is crucial to ensure action invariance under diff.s.
  
 Can we generalize form integration to {\it supermanifolds}, and use it to construct actions 
 automatically invariant under superdiffeomorphisms ?  The answer to both questions is affirmative.
 
 In analogy 
 with the bosonic case, integration on forms living on supermanifolds is defined via integration of 
functions in superspace. Consider a function $\Phi (x, \theta)$, defined on a supermanifold $M^{d|m}$ 
with $d$ bosonic coordinates $x$ 
and $m$ fermionic (anticommuting) coordinates $\theta^\al$. It is called a {\it superfield}, and can be expanded in the $\theta^\al$ coordinates:
\eq
\Phi(x,\theta)= \phi(x) + \phi_{\alpha_1} (x) \theta^{\alpha_1}+ \phi_{\alpha_1\alpha_2} (x) \theta^{\alpha_1} \theta^{\alpha_2} + \cdots + \phi_{\alpha_1 \cdots \alpha_m} (x) \theta^{\alpha_1} \cdots \theta^{\alpha_m} \label{superfield}
\en
The functions $\phi_{{\al_1} \cdots {\al_p}}(x)$ are called {\it superfield components} ,
and have antisymmetrized indices due to the anticommuting $\theta$'s in the expansion
 (\ref{superfield}). The integral of
the superfield on $M^{d|m}$ is defined by Berezin integration: 
\eq
\int_{M^{d|m}} \Phi (x,\theta) ~d^dx ~d^m \theta \equiv \int_{M^d} {1 \over m!} \phi_{\alpha_1 \cdots \alpha_m} (x)  \epsi^{\alpha_1 \cdots \alpha_m}~ d^dx  \label{Berezin}
\en
Only the highest component of $\Phi$ (corresponding to the maximal number of $\theta$'s) enters the integral on $M^d$.
 
 Note the striking similarity between the two integrals (\ref{intform}) and (\ref{Berezin}). In fact
we can define form integration in terms of Berezin integration. Consider the
 differentials $dx$ in the $d$-form (\ref{dform}) as {\it anticommuting coordinates} $\xi^\mu = dx^\mu$,
 so that $\om^{(d)}$ becomes a {\it function} of $x$ and $\xi$:
 \eq
 \omega^{(d)} (x,\xi) =\omega_{[\mu_1 \cdots \mu_d]} (x) ~\xi^{\mu_1} \cdots  \xi^{\mu_d} 
 \en
Its Berezin integral on $M^{d|d}$ exactly yields the form integral (\ref{intform}). This observation
 is the key for a definition of superform integration on supermanifolds. 
 
A natural generalization of a bosonic top form (\ref{dform}) is a $(d+m)$-superform:
  \eq
  \omega^{(d+m)} (x, \theta)= \omega_{[\mu_1 \cdots \mu_d] \{\alpha_1 \cdots \alpha_m \} } (x, \theta)~
 dx^{\mu_1} \wedge \cdots \wedge dx^{\mu_d} \wedge d\theta^{\alpha_1} \wedge \cdots \wedge d \theta^{\alpha_m}
\label{superform}
 \en
 
\noi  Note that the $d\theta$ differentials {\it commute} (since the $\theta$'s are anticommuting),
  so that the indices $\alpha_i$ are symmetrized. For this reason $\omega^{(d+m)}$ cannot be a top form: a superform can have an arbitrary number of $d \theta$ differentials, and  its exterior derivative does not vanish. Let's ignore for the moment this difficulty, and try to define a superform integral. Inspired by 
the observation in the preceding paragraph, we consider the superform $\omega^{(d+m)}(x,\theta)$
as a function of $x,\theta,dx,d\theta$, i.e. a function of the commuting variables $x, d\theta$ and
anticommuting variables $\theta, dx$. Its integral can be defined by Berezin integration on
 $\theta,dx$, and usual Riemann-Lebesgue integration on $x,d\theta$. Here a second difficulty arises:
 the ordinary integration on the $u= d\theta$ coordinates produces integrals of the type 
 \eq
 \int u^m ~d^m u
 \en
and there is no algorithmic way to assign a $C$-number to it.
For the integral on the even variables $u=d\theta$ to make sense, the integrand must have compact support as a function of $u$.  For this reason we consider functions of the $d\theta$'s which are {\it distributions} in $d\theta$ with support at the origin:
\eq
\omega (x,\theta,dx,d\theta)= \omega_{[\mu_1 \cdots \mu_d]} (x,\theta)~ dx^{\mu_1} \cdots dx^{\mu_d} \delta(d\theta^{1}) \cdots \delta(d\theta^{m})
\en
These ``functions" can be integrated on the supermanifold $M^{d+m|d+m}$
spanned by the $d+m$ bosonic variables $x,d\theta$ and $d+m$ fermionic variables $dx,\theta$.
The integral
\eq
\int_{M^{d+m|d+m}} \omega (x,\theta,dx,d\theta) ~d^dx~ d^m\theta ~d^d(dx) ~d^m (d\theta)
\en
is defined by Berezin integration on the odd variables $dx,\theta$ and usual Riemann-Lebesgue 
integration on the even variables $x,d\theta$. Carrying out integration on the variables $dx$ and $d\theta$ the
integral becomes
\eq
\int_{M^{d|m}} \omega_{[\mu_1 \cdots \mu_d]} (x,\theta) \epsilon^{\mu_1 \cdots \mu_d} ~d^dx ~d^m \theta
\label{intdm}
\en
 This integral can also be seen as
an integral of the {\it form}:
\eq
\omega^{d|m} = \omega_{[\mu_1 \cdots \mu_d]} (x,\theta)
\delta(u^{1}) \cdots \delta(u^{m})~ dx^{\mu_1} \we \cdots \we dx^{\mu_d} \we du^{1} \we\cdots \we du^{m} \label{inttop}
\en
where the even variables $u$ are the differentials $d\theta$.  Indeed, let us integrate this form with 
the recipe of considering it a function of $x,\theta,u$ and of the differentials $dx$,$du$, and then using Berezin and Riemann integration according to the odd or even grading of the variables. The result coincides with  (\ref{intdm}).

Thus the form $\omega^{d|m}$ can be integrated, even if it contains $d\theta$ differentials. 
We achieve this by confining the $d\theta$'s inside delta functions, and in this way overcome the first difficulty 
encountered with the superforms (\ref{superform}). But can $\omega^{d|m}$ overcome also the
second difficulty, and be a {\it top form}? The answer is yes: the $dx$ and $du$ differentials are all anticommuting, so that their number
in $\om^{d|m}$ is already maximal, and multiplying it by $d\theta$ differentials gives zero because
of the presence of the deltas. Therefore $d\om^{d|m}=0$, and $\om^{d|m}$ is a {\it bona fide} top form.
Since it can be integrated and it is a top form, $\om^{d|m}$ is called an {\it integral top form}.

Finally, using the notation
\eq
\delta(u^{1}) \cdots \delta(u^{m})~ du^{1} \we\cdots \we du^{m} \equiv \delta(u^{1}) \we \cdots \we \delta(u^{m})~ 
\en
the integral top form can be rewritten (using $u=d\theta$):
\eq
\omega^{d|m} = \omega_{[\mu_1 \cdots \mu_d]} (x,\theta)
~ dx^{\mu_1} \we \cdots \we dx^{\mu_d} \we \delta(d\theta^{1}) \we \cdots \we \delta(d\theta^{m}) \label{inttop2}
\en
or also
\eq
\omega^{d|m} = \omega_{[\mu_1 \cdots \mu_d] [\al_1 \cdots \al_m]} (x,\theta)
~ dx^{\mu_1} \we \cdots \we dx^{\mu_d} \we \delta(d\theta^{\al_1}) \we \cdots \we \delta(d\theta^{\al_m}) \label{inttop3}
\en
where indices $\al$ are antisymmetrized since the $\delta (d\theta^\al)$ anticommute, and
\eq
m! ~\omega_{[\mu_1 \cdots \mu_d]} \equiv \omega_{[\mu_1 \cdots \mu_d] [1 \cdots m]}
\en
In this notation $\mu$ and $\al$ indices play a similar role, and are both antisymmetrized. The numbers
$d,m$ are respectively called the {\it form number} and the {\it picture number}, and for integral top forms
they coincide with the numbers of bosonic and fermionic dimensions of the supermanifold $M^{d|m}$.

We call ``superforms" the forms of the kind (\ref{superform}), with $dx$ and $d\theta$ differentials, 
without $\delta (d\theta)$'s. Thus superforms have a form number that counts the $dx,d\theta$
differentials, and zero picture number. For example the Lagrangian in (\ref{GintegralSP}) is a 
superform $L^{4|0}$. 

\sk
\sk
\noi {\bf Integration on submanifolds of supermanifolds}
\sk
Supergravity actions on supergroup manifolds $\Gtilde$ are given by integrals of a $d$-form Lagrangian $L$ on
a $d$-dimensional (bosonic) submanifold $M^d$ of $\Gtilde$. They can be written as integrals on the whole
$\Gtilde$ of the Lagrangian multiplied by an appropriate Poincar\'e dual $\eta_{M^d}$ of $M^d$, such that
$L \we \eta_{M^d}$ becomes an integral top form. Let us see how this works for $N=1$, $d=4$ supergravity.

The supergravity Lagrangian in (\ref{GintegralSP}) is a $(4|0)$ superform. For simplicity we now assume
that fields satisfy the Lorentz horizontality constraints on all the curvatures, and thus effectively
depend only on the superspace coordinates $x^\mu, \theta^\alpha$, with $\mu=1,..4$, $\al=1,..4$.
Then $\Gtilde$ is $M^{4|4}$ superspace, and only integral top forms of type $(4|4)$ can be 
integrated on $M^{4|4}$. We therefore need a Poincar\'e dual of type $(0|4)$, so that
\eq
L^{4|0} \we \eta_{M^4}^{0|4} 
\en
is an integral top form, i.e. of type $(4|4)$. For this purpose we choose:
\eq
\eta_{M^4}^{0|4} = \epsi_{\al\be\ga\de} \theta^\al \theta^\be \theta^\ga \theta^\de ~ \epsi_{\al'\be'\ga'\de'}
\delta (d\theta^{\al'}) \we \delta (d\theta^{\be'}) \we \delta (d\theta^{\ga'}) \we \delta (d\theta^{\de'})
\en
so that
\eq
\int_{M^{4|4}} L^{4|0} \we  \eta_{M^4}^{0|4} = \int_{M^{4}} L^{4|0} (\theta=0, d\theta=0)
\en
and we obtain a spacetime action, where all fields depend only on $x$-coordinates (the terms containing 
$\theta$'s are annihilated by the presence of the 4 $\theta$'s in $\eta$) and have no ``legs"  
$d\theta$ because of the $\delta (d\theta)$ in $\eta$. Note that $\eta_{M^4}^{0|4}$ is closed,
and the explicit $\theta$'s prevent it to be exact.

Since multiplying by the Poincar\'e dual changes the picture number of the resulting form,
$\eta$ is also called Picture Changing Operator (PCO), a name borrowed from string theory
and string field theory.

The Poincar\'e dual is by no means unique: we can orient the $M^4$ surface inside $\Gtilde$
in many different ways. For example consider the PCO obtained by acting on $\eta$ with an infinitesimal diffeomorphism in the $\theta$ directions:
\eq
\eta ' = \eta + \ell_\epsilon \eta = \eta + d(\iota_\epsilon \eta)  \label{etachange}
\en
This is still a PCO, being closed and not exact\footnote{because $\eta$ is closed and not exact, and 
$d$ commutes with $\ell_\epsilon$.}, and dual to a submanifold diffeomorphic to the original $M^4$.
Note also that the change in $\eta$ is exact.

 \sect{Gamma matrices in $d=3+1$}
 
\eqa
& & \eta_{ab} =(1,-1,-1,-1),~~~\{\ga_a,\ga_b\}=2 \eta_{ab},~~~[\ga_a,\ga_b]=2 \ga_{ab}, \\
& & \ga_5 \equiv  -i \ga_0\ga_1\ga_2\ga_3,~~~\ga_5 \ga_5 = 1,~~~\epsi_{0123} = - \epsi^{0123}=1, \\
& & \ga_a^\dagger = \ga_0 \ga_a \ga_0, ~~~\ga_5^\dagger = \ga_5 \\
& & \ga_a^T = - C \ga_a C^{-1},~~~\ga_5^T = C \ga_5 C^{-1}, ~~~C^2 =-1,~~~C^T =-C
\ena

\subsection{Useful identities}
\eqa
 & &\ga_a\ga_b= \ga_{ab}+\eta_{ab}\\
 & & \ga_{ab} \ga_5 = - {i \over 2} \epsilon_{abcd} \ga^{cd}\\
 & &\ga_{ab} \ga_c=\eta_{bc} \ga_a - \eta_{ac} \ga_b +i \epsi_{abcd}\ga_5 \ga^d\\
 & &\ga_c \ga_{ab} = \eta_{ac} \ga_b - \eta_{bc} \ga_a +i \epsi_{abcd}\ga_5 \ga^d\\
 & &\ga_a\ga_b\ga_c= \eta_{ab}\ga_c + \eta_{bc} \ga_a - \eta_{ac} \ga_b +i \epsi_{abcd}\ga_5 \ga^d\\
 & &\ga^{ab} \ga_{cd} = i \epsi^{ab}_{~~cd}\ga_5 - 4 \de^{[a}_{[c} \ga^{b]}_{~~d]} - 2 \de^{ab}_{cd}
 \ena

 \subsection{Charge conjugation and Majorana condition}

\eqa
 & &   {\rm Dirac~ conjugate~~} \psibar \equiv \psi^\dagger
 \ga_0\\
 & &  {\rm Charge~ conjugate~spinor~~} \psi^c = C (\psibar)^T  \\
 & & {\rm Majorana~ spinor~~} \psi^c = \psi~~\Rightarrow \psibar =
 \psi^T C
 \ena

\subsection{Fierz identity for two spinor one-forms}
\eq
 \psi  \chibar = \unquarto [ (\chibar  \psi) 1 + (\chibar \ga_5  \psi) \ga_5 + (\chibar \ga^a  \psi) \ga_a + (\chibar \ga^a \ga_5  \psi) \ga_a \ga_5  - \unmezzo (\chibar \ga^{ab}  \psi) \ga_{ab}]
 \en
 \subsection{Fierz identity for two Majorana spinor one-forms}
 \eq
 \psi  \psibar = \unquarto [  (\psibar \ga^a  \psi) \ga_a  - \unmezzo (\psibar \ga^{ab}  \psi) \ga_{ab}]
 \en
 \noi As a consequence
 \eq
\ga_a \psi \psibar \ga^a \psi =0,~~~ \psi \psibar \ga^a \psi- \ga_b \psi \psibar \ga^{ab} \psi=0 \label{Fierz4d}
\en

\vfill\eject
\end{document}